\documentclass[english]{article}
\usepackage[T1]{fontenc}
\usepackage[latin9]{inputenc}
\usepackage{amsmath}
\usepackage{amsthm}
\usepackage{amssymb}
\usepackage{graphicx}

\makeatletter
\numberwithin{figure}{section}
\newcommand{\lyxaddress}[1]{
\par {\raggedright #1
\vspace{1.4em}
\noindent\par}
}

\makeatother

\usepackage{babel}
\begin{document}

\title{\textbf{Hartle-Hawking boundary conditions as Nucleation by de Sitter Vacuum}}

\author{\textbf{$^{1}$F. Feleppa, $^{2}$I. Licata and $^{2}$C. Corda}}
\maketitle

\lyxaddress{\textbf{$^{1}$Department of Physics, University of Trieste, via
Valerio 2, 34127 Trieste, Italy}}

\lyxaddress{\textbf{$^{2}$Research Institute for Astronomy and Astrophysics
of Maragha (RIAAM), P.O. Box 55134-441, Maragha, Iran and International
Institute for Applicable Mathematics and Information Sciences, B.
M. Birla Science Centre, Adarshnagar, Hyderabad 500063, India}}
\begin{abstract}
\raggedright It is shown that, for a de Sitter Universe, the  Hartle-Hawking
(HH) wave function can be obtained in a simple way starting from the Friedmann-Lemaitre-Robertson-Walker (FLRW) line element of  cosmological equations. An oscillator having imaginary  time is indeed derived starting from the Hamiltonian obtaining the HH condition. This proposes again some crucial matter on the meaning of complex time in cosmology. In order to overcome such difficulties, we propose an interpretation  of the HH framework based on de Sitter Projective Holography.
\end{abstract}
\begin{description}
\item [{PhySH:}] Gravitation, Cosmology \& Astrophysics/Cosmology/Quantum
Cosmology.
\item [{Keywords:}] Hartle-Hawking wave function; de Sitter Projective
Holography; complex time; quantum oscillator.
\end{description}

\section{Introduction}

The Hartle-Hawking (HH) proposal of ``\emph{no boundary conditions}''
\cite{key-1} is, till now, the most powerful and fruitful theory
on the original quantum state of the Universe. Its original motivation
was to avoid the initial singularity of the Universe due to the cosmological
application of the general theory of relativity (GTR) \cite{key-10}.
The HH proposal concerns the state of the Universe before the Planck
time, that is, before the emerging of the classical space-time by
an overlapping of quantum patterns \cite{key-1}. Such an emerging
is realized through a functional on the geometries of compact three-manifolds
and on the values of the matter-fields on such manifolds \cite{key-1}.
In that way, the wave function of the Universe's ground state is obtained
by a path integral over all compact positive-definite four-geometries
having the three-geometry as a boundary \cite{key-1}. It implies
a constraint on the Hermitian behavior of the Hamiltonian which satisfies
the famous Wheeler-DeWitt wave equation for the Universe and enables
its computability \cite{key-11}. After the initial HH proposal, the
no boundary conditions have been discussed in other, different scenarios
like multiverse, brane world, inflation, quantum potential {[}12-15{]}.
In such works, it is assumed that the ``vacuum'' before the Big-Bang
represents the absence of a classical space-time. An interesting question
concerns if such an assumption of absence of a classical space-time
works beyond cosmology, maybe including the whole realm of quantum
theory \cite{key-16}. Some recent foundational scenarios go in the
same direction \cite{key-17,key-18}. In this work, the HH proposal
is reanalysed starting from its recent generalization in \cite{key-19}.
In Section 2, a way to construct the Lagrangian starting from the
Friedmann-Lemaitre-Robertson-Walker (FLRW) line element is considered
and an oscillator having imaginary time is derived starting from the
Hamiltonian obtaining the HH condition. In Section 3 we discuss some
problem of de Sitter geometry. In Section 4 we focus on the meaning
of the derived oscillator within a de Sitter framework with a bird
view vision on de Sitter Projective Cosmology. Finally, we will conclude
with a discussion concerning time and wave function in cosmology. 

\section{\large A Cosmological Lagrangian for the Hartle-Hawking conditions}

In the recent work \cite{key-19}, Hartle and collaborators reconsidered
the nature of the no-boundary conditions. In that way, they proposed
a generalization which should facilitate the understanding of the
HH proposal and its application to observational studies of the Universe.
The basic idea in \cite{key-19} is that the HH proposal arises from
a semi-classical constraint which results completely independent by
specific assumptions on the real nature of the quantum theory of gravity.
This is different with respect to the pioneering papers by Hawking
and collaborators where the path integrals approach has been applied
to quantum fluctuations, see for example \cite{key-20,key-21}. In
fact, in such works the connection between path integrals and euclidean
geometry was not only considered in terms of a constructive approach,
but also as a possible connection with quantum gravity based on its
specific application to black hole radiance. This generated various
problems also in the semi-classical approach to the no-boundary wave
function (NBWF) through tunnelling effect, which was been introduced
by Vilenkin \cite{key-22,key-23}. In a remarkable paper \cite{key-24},
Lehners and Turok has shown that, for a large class of theories, a
semi-classical approach to the NBWF through path integrals is incompatible
with the Euclidean approach. The Euclidean issue, which will be partially
discussed also in the present paper, is very controversial, see for
example \cite{key-25}.

In \cite{key-19}, the NBWF has been assumed to be the Universe's ground
state and the authors have shown that path integrals are not needed
if one uses general assumptions. The most important among such assumptions
are the compatibility between the GTR and quantum mechanics (QM) and
the definition of the wave function by a collections of saddle points
which guarantee the opportune smoothness between the quantum ground
state and the subsequent evolution of the space-time. 

In the present work, it is chosen to construct the wave function starting
from the Universe's scale factor through a quantum-like approach.
After that, it will be shown that such a wave function satisfies the
HH no-boundary condition.

The starting point is to construct a Lagrangian from the scale factor
of the Universe, which gives the FLWR equations \cite{key-3}. Let
us consider the following Lagrangian for a homogeneous and isotropic
Universe:

\begin{equation}
L(R,\dot{R})=\dot{R}^{2}+\frac{8\pi G\rho}{3}R^{2}+\frac{\Lambda c^{2}}{3}R^{2},\label{eq: 1}
\end{equation}
where $R=R(t)$, $\rho=\rho(R)$ is the energy density of matter in
the Universe, and $\Lambda$ is the cosmological constant introduced
by Einstein in 1917 \cite{key-4}. Starting from Eq. (\ref{eq: 1}),
one derives the first FLRW equation. One considers the scale factor
$R$ as being a generalized coordinate. Then, the energy of the Universe
is: 
\begin{equation}
E(R,\dot{R})=\left(\frac{\partial L}{\partial\dot{R}}\right)\dot{R}-L=\dot{R}^{2}-\frac{8\pi G\rho}{3}R^{2}-\frac{\Lambda c^{2}}{3}R^{2}.\label{eq: 2}
\end{equation}
This energy is constant because the Lagrangian does not depend explicitly
on time \cite{key-5}. Denoting the constant $E$ by $-kc^{2}$, one
obtains

\begin{equation}
\left(\frac{\dot{R}}{R}\right)^{2}+\frac{kc^{2}}{R^{2}}=\frac{8\pi G\rho}{3}+\frac{\Lambda c^{2}}{3}.\label{eq: 3}
\end{equation}
This equation is the first equation for the standard Lambda-CDM model
\cite{key-6}, which can be closed ($k=1$), flat ($k=0$), or open
($k=-1$) \cite{key-3}. In order to get a unit of energy, one multiplies
the Lagrangian (\ref{eq: 1}) by a constant mass:

\begin{equation}
L(R,\dot{R})=m\left(\dot{R}^{2}+\frac{8\pi G\rho}{3}R^{2}+\frac{\Lambda c^{2}}{3}R^{2}\right).\label{eq: 4}
\end{equation}
From the above relations the Hamiltonian can be written down as 

\begin{equation}
H(R,\dot{R})=\frac{1}{2}m\dot{R}^{2}-\frac{4\pi mG\rho}{3}R^{2}-\frac{\Lambda mc^{2}}{6}R^{2}.\label{eq: 5}
\end{equation}
We consider a pure de Sitter Universe, which simplifies the Lambda-CDM
Cosmology by modelling the Universe as spatially flat and neglects
ordinary matter \cite{key-7}. Thus, the dynamics of the Universe
are dominated by the cosmological constant, which should correspond
to Dark Energy in the present Universe or to the inflaton field in
the early Universe. In this case, the first Friedmann equation becomes:

\begin{equation}
\left(\frac{\dot{R}}{R}\right)^{2}=\frac{\Lambda c^{2}}{3},\label{eq: 6}
\end{equation}
and the Hamiltonian

\begin{equation}
H(R,\dot{R})=\frac{1}{2}m\dot{R}^{2}-\frac{\Lambda mc^{2}}{6}R^{2}.\label{eq: 7}
\end{equation}
Following \cite{key-56}, one constructs a phase space $(R,P_{R})$, where $R$ is the
comoving coordinate with

\begin{equation}
P_{R}=m\dot{R}\label{eq: 8}
\end{equation}
being the linear momentum. Thus, inserting Eq. (\ref{eq: 8}) into
Eq. (\ref{eq: 7}), one obtains

\begin{equation}
H(R,\dot{R})=\frac{P_{R}^{2}}{2m}-\frac{\Lambda mc^{2}}{6}R^{2}.\label{eq: 9}
\end{equation}
Therefore, one goes ahead with the canonical quantization \cite{key-8}

\begin{equation}
P_{R}\rightarrow\mathcal{P_{R}}=-i\hbar\frac{\partial}{\partial R},\label{eq: 10}
\end{equation}
obtaining the Hamiltonian operator

\begin{equation}
\mathcal{H}=-\frac{\hbar^{2}}{2m}\frac{\partial^{2}}{\partial R^{2}}-\frac{\Lambda mc^{2}}{6}R^{2}.\label{eq: 11}
\end{equation}
Rescaling $\Lambda$ and setting $\hbar=c=m=1,$ one gets

\begin{equation}
\mathcal{H}=\frac{1}{2}(\mathcal{P_{R}}^{2}-\Lambda R^{2}).\label{eq: 12}
\end{equation}
which is exactly the Hamiltonian of a simple harmonic oscillator.
The frequency is determined by $\sqrt{\Lambda}$. At this point, one
considers three cases of interest: $\Lambda=0$ (free particle), $\Lambda<0$
(oscillator) and $\Lambda>0$ (inverted oscillator). Firstly, let
us consider the case $R\in(0,\infty)$. The obvious choice for the
Hilbert space is $L^{2}(\mathbb{R^{+}},dR)$. In this space the Hamiltonians
of the form $\mathcal{P_{R}}^{2}+V(R)$ have self-adjoint extensions
\cite{key-8}. Specifically, it is readily checked that the physical
Hamiltonian (\ref{eq: 12}) is symmetric in the usual representation
$\mathcal{P_{R}}^{2}\rightarrow-i\frac{\partial}{\partial R}$, provided
$\lim_{x\rightarrow\infty}\phi=0$ and $\lim_{x\rightarrow0}[\psi^{*}\phi-\phi\psi^{*}]=0$.
This gives the boundary condition $\phi'(0)=\alpha\phi(0),\alpha\in\mathbb{R}$.
The Hilbert space is the subspace specified by

\begin{equation}
\mathbb{H_{\alpha}}=\left\{ \phi\in L^{2}(\mathbb{R^{+}},dR)\right\} .\label{eq: 13}
\end{equation}
Following \cite{key-9}, one solves the time-dependent Schr\"odinger
equation

\begin{equation}
i\frac{\partial}{\partial t}\phi(R,t)=-\frac{1}{2}\frac{\partial^{2}}{\partial R^{2}}\phi(R,t)-\frac{1}{2}\Lambda R^{2}\phi(R,t),\label{eq: 14}
\end{equation}
with the above mentioned boundary condition. 

If $\Lambda=0$, there are two types of elementary solutions: ingoing
and outgoing waves of fixed energy and a bound state. The first one,
satisfying the above boundary condition, has the form

\begin{equation}
\phi_{\alpha k}(R,t)=e^{ik^{2}t/2}\left[e^{ikR}-\left(\frac{\alpha-ik}{\alpha+ik}\right)e^{-ikR}\right].\label{eq: 15}
\end{equation}
Concerning the second type of solution, a bound state, one gets 

\begin{equation}
\phi(R,t)=e^{i\textmd{k}^{2}t/2}e^{-\textmd{k}R},\hspace{0.5cm}\textmd{k}>0.\label{eq: 16}
\end{equation}
In the case of $\Lambda<0$ (AdS Universe), one has the oscillator
on the half-line with the usual boundary condition. With $\Lambda=-1/l^{2}$
and $\zeta=t/l$, the propagator on $\mathbb{R}$ is

\begin{equation}
K(R,\zeta,R',0)=\sqrt{\frac{1}{2\pi il\sin(\zeta)}}\exp\left\{ \frac{i[(R^{2}+R'^{2})\cos(\zeta)-2RR']}{2l\sin(\zeta)}\right\}. \label{eq: 17}
\end{equation}
Now, let us consider the region $R<0$. Given $\psi(R,0)=f(R)$ for
$x>0$, the initial data $f(R)$ on $\mathbb{R^{+}}$ can be extended
to the region $R<0$, such that

\begin{equation}
f'(R)-\alpha f(R)=-(f'(-R)-\alpha f(-R)),\hspace{0.5cm}R<0,\label{eq: 18}
\end{equation}
that is, imposing antisymmetry on the boundary condition function.
If one solves this equation, one obtains the required extension

\begin{equation}
f_{L}(R)\equiv e^{\alpha R}\int_{R}^{0}e^{-\alpha u}f'(-u)-\alpha f(-u)\,du+e^{\alpha R}f(0),\hspace{0.5cm}R<0,\label{eq: 19}
\end{equation}
where the integration constant is chosen such that $f_{L}(0)=f(0)$.
Convoluting the data so extended with the full-line propagator (\ref{eq: 17}),
one gets the solution

\begin{equation}
\psi(R,\zeta)=\int_{-\infty}^{0}K(R,\zeta,R',0)f_{L}(R')\,dR'+\int_{0}^{\infty}K(R,\zeta,R',0)f(R')\,dR',\hspace{0.5cm}R>0.\label{eq: 20}
\end{equation}
If $\Lambda<0$, the Hamiltonian is not bounded below. However the
unitary evolution operator is still well defined since the Hamiltonian
has self-adjoint extensions. The propagator on $\mathbb{R}$ is obtained
by the replacement $l\rightarrow il$ to give

\begin{equation}
\bar{K}(R,\zeta,R',0)=\sqrt{\frac{1}{2\pi il\sinh(\zeta)}}\exp\left\{ \frac{i[(R^{2}+R'^{2})\cosh(\zeta)-2RR']}{2l\sinh(\zeta)}\right\} .\label{eq: 21}
\end{equation}
Starting from this expression, setting $R'=0$, one obtains the HH
wave function:

\begin{equation}
\Psi_{HH}\equiv\bar{K}(R,\zeta,0,0)=\sqrt{\frac{1}{2\pi il\sinh(\zeta)}}\exp\left\{ -\frac{iR^{2}}{2l\tanh(\zeta)}\right\} .\label{eq: 22}
\end{equation}

It is a remarkable issue that, starting from a series of simple and
quantum-like assumptions, one can get the HH wave function, with the
important interpretative contribution of the harmonic oscillator of Eq. 
(\ref{eq: 12}) (another approach can be found in \cite{key-9}).
Thus, our analysis seems to confirm the well known statement by Sidney
Coleman that \textquotedbl{}The career of a young theoretical physicist
consists of treating the harmonic oscillator in ever-increasing levels
of abstraction\textquotedbl{} \cite{key-26}. The issue that one can
obtain HH from Eq. (\ref{eq: 12}) is consistent with the idea of
Vilenkin on the presence of a tunnelling effect \cite{key-22,key-23}.
In addition, one observes that the frequency of the oscillator depends
on the cosmological constant. This can suggest a special role of $\Lambda$,
which goes well beyond the large-scale geometrical structure. We will
further discuss this special role of $\Lambda$ in the following of
the paper.

\section{Problems with de Sitter geometry}

Now, let us ask if the physical interpretation of the HH function
can be considered exhaustive or, alternatively, if we still need some
further conceptual tessera. In fact, the interpretation of QM in a
cosmological framework needs particular rules. It is sufficient recalling
that the whole array of matters connected to the role of the observer
(like preparing a quantum state and its measures) does not work in
quantum cosmology. This is efficaciously stressed by R. Serber \textquotedbl{}Observers
were not present at the Big-Bang instant\textquotedbl{} \cite{key-27}.
This generated an interesting, hot debate with various opposing opinions,
on one hand, on the cosmological perspectives of different approaches
(de Broglie-Bohm theory, Decoherent Histories Approach to Quantum
Mechanics, Everettian Many Worlds); on the other hand on the role
of some fundamental rule of ``quantum data'', like the Born Rule
\cite{key-28,key-29}. A very important issue is that, generally speaking,
QM, with its various interpretations, refers to a space-time where
the wave function evolves. On the contrary, in quantum cosmology the
wave function \emph{generates} the same space-time. In fact, the origin
of quantum cosmology basically arises from attempts to remove the
classical initial singularity at the quantum level and from a starting
point to study the \textquotedbl{}subtle fabric\textquotedbl{} of
quantum gravity. Thus, researchers have a general agreement only on
few issues, but such few issues are sufficient in order to understand
that, beyond its formal fineness, quantum cosmology is still at an
early stage where one has no definitive interpretations because the
old problem of the ``Peaceful Coexistence'' between the Classical
Space-Time and the Quantum Realm is still present and very strong
\cite{key-30}. 

Concerning our discussion on the HH wave function, a suggestive way
to describe the physical situation is telling that one finds a ``quantum
nebula'' which replaces the classical initial singularity. This ``quantum
nebula'' generates an accelerate expansion of the space-time structure.
This has not to be considered \textquotedbl{}a collapse\textquotedbl{}.
Instead, it is a kind of natural smooth transition which is completely
unprecedented in QM. In addition, one must consider a metric structure
as being the description of a physical field on a fixed background
(usually, that background has constant curvature) \cite{key-31}.
Hence, the main task of the HH wave function is to produce the collapse
of a very high number of freedom degrees of quanta through a kind
of ``nucleation''. In other words, a space-time which is described
by a differentiable manifold is an emergent phenomenon while the metric
and the Christoffell connections are collective variables which can
work only at lows energies and long wavelengths. This approach concerns
some recent theories of superfluid vacuum \cite{key-32,key-33}, while,
in the recent work \cite{key-18}, a unitary approach to both the
particle-like \textquotedbl{}Little Bangs\textquotedbl{} and the cosmological
\textquotedbl{}Big-One\textquotedbl{} where the wave function is always
interpreted as an event-based emergent phenomenon has been proposed.
The oscillator (\ref{eq: 12}) seems a natural answer to the idea
of nucleation and a potential strong connection with the value of
the cosmological constant appears to be a logical consequence of the
approach of this paper. Thus, problems concerning the physical interpretation
of the HH wave function are transferred to the oscillator (\ref{eq: 12})
which works with imaginary time. Actually, this issue revealed itself
in the original derivation of the HH wave function \cite{key-1}.
The ``quantum nebula'' is indeed technically described by Hartle
\& Hawking in terms of a de Sitter-like Universe having imaginary
time. In this framework, space and time are indistinguishable and
close to each other. Hawking and collaborators stressed in various
works (see for example \cite{key-34}) that the subsequent parallel
sections of that hypersphere should represent a hyperspherical, three-dimensional
space which starts from a non-singular ``South Pole''. This pole
expands as far as a maximum dimension and, after that, it reduces
to a ``North Pole'', which is still non-singular as the global space
has a hyperspherical behavior. This idea came first the cosmological
evidence of an accelerating Universe with the triumphant comeback
of the cosmological constant, but it represents the essence of the
line of reasoning of Hawking and collaborators. The problem is that
such a line of reasoning seems erroneous. In fact, if one assumes
that imaginary time is portrayed by a maximum circle, the spatial
sections, which are perpendicular to the points of the circle, are
S3 space having constant section! Then, in the absence of a ``collapse''
(or some exotic mechanism), one cannot understand in which way a de
Sitter Universe having \emph{imaginary} time becomes a FLRW Universe
having \emph{real} time. On the other hand, it is \emph{exactly} a
de Sitter hyperspherical Universe having imaginary time, with its new,
central position in cosmology, which tells one that the HH approach
is an intriguing idea which opens to various interesting possibilities.
Here, we propose an alternative method in order to analyse de Sitter
Universe, which is based on the approach of the so called de Sitter
Projective Relativity \cite{key-35}. We think that in this scenario
the HH wave function and the oscillator (\ref{eq: 12}) can find a
new, enlightening interpretation. 

\section{The Physical Meaning of Complex Time in de Sitter Projective Holography}

The Theory of de Sitter Projective Holography was born within a program
of Quantum Cosmology and has also developed a new approach to Particle
Physics {[}36 - 40{]}. For our purpose, it suffices here to define
Vacuum as a Universal Action Reservoir \cite{key-41} placed on a
4D surface of an Euclidean five-dimensional (5D) hypersphere. This
surface can be converted into a 4D hyperboloid which represents a
de Sitter space\textendash time by a Wick rotation. The Beltrami projective
representation of this de Sitter space\textendash time on a 4D hyperplane
tangent to the hyperboloid in the point-event of observation is known
as \textquotedblleft Castelnuovo chronotope.\textquotedblright{} It
is important to remark that the group approach \cite{key-42,key-43}
makes possible to individuate the de Sitter Universe as a framework
of physical processes without referring to any \textquotedblleft local\textquotedblright{}
physics. The de Sitter group is in fact the simplest 4D group containing
those of Galilei and Lorentz-Poincar\'e. We note that it is not necessary
to imagine the hypersphere as an \textquotedblleft enlarged\textquotedblright{}
space-time. The space-time labels of observable
events do not belong to the archaic phase, represented on the hypersphere:
it is populated with virtual processes only. We called such theory
\textquotedblleft Archaic\textquotedblright{} because the role of
the hypersphere and that of the observed Universe are not related
by a \textquotedblleft before\textquotedblright{} and an \textquotedblleft after\textquotedblright .
The hypersphere is rather like a highly non-local, a-temporal phase.
The space-time positions or \textquotedblleft labels\textquotedblright{}
are, instead, related to the descriptions of the classes of observers
on the tangent $P^{4}$-plane (Beltrami-Castelnuovo Projective representation,
see figure 4.1). 
\begin{figure}

\includegraphics[scale=0.1]{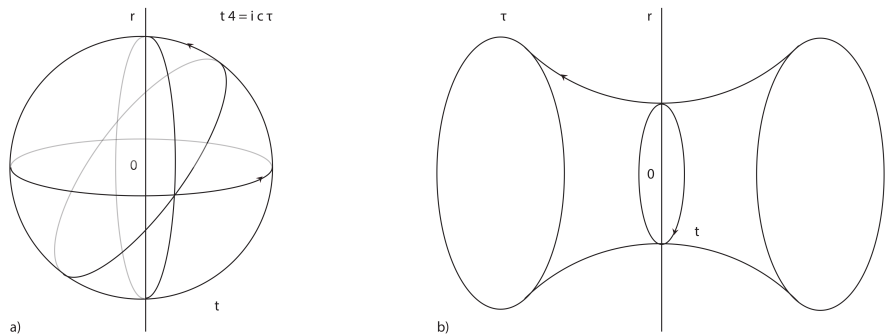}\caption{Beltrami-Castelnuovo Projective representation: a) imaginary time and b) real time.}

\end{figure}
 The passage from a description to the other one is defined by the
Wick rotation \cite{key-44}: before jumping out from vacuum, particles
are in a virtual status described by the imaginary time of pre-space;
during the observable existence lag, the real time comes into play.
The manifestation of particles from the vacuum and their disappearance
into the vacuum are the real microinteractions described in quantum
theory through the wave-function ``collapse'' or reduction (R processes).
To all effects, a ``localization'' at the level of individual event
in microphysics, a ``nucleation'' in the case of the Big Bang. Such
strong unity between macro and microphysics justifies the consideration
of an archaic holography ruled by a Wick rotation and projectivity.
Now, let us take a closer look at the role of imaginary time. We can
think of an axis $\underline{x}_{0}$ of the hypersphere representing
inverse temperatures, and imagine that below a critical value of this
coordinate the physical processes are constrained to remain virtual.
This constraint is removed when $\underline{x}_{0}$ exceeds the critical
value, thus permitting the emergence of real processes in real time,
at a rate completely analogous to the exponential one of the radioactive
decay. This axis can be considered as an \textquotedblleft archaic
precursor\textquotedblright{} of time. The 5-sphere is: 
\begin{equation}
(\underline{x}_{0})^{2}+(\underline{x}_{1})^{2}+(\underline{x}_{2})^{2}+(\underline{x}_{3})^{2}+(\underline{x}_{5})^{2}=r^{2}.\label{eq: 23}
\end{equation}
It is reasonable to assume the critical value as corresponding to
baryogenesis temperature: 
\begin{equation}
T_{C}=\frac{\hbar}{k\theta_{0}},\label{eq: 24}
\end{equation}
where $\theta_{0}\sim10^{-23}$ s plays the role of the fundamental
time interval (chronon), so $T_{C}\approx10^{13}$ K. We will come
back later on the chronon and its relations with
timescales. One can believe, without too much effort, that even in
the archaic phase the state of matter could still be described by
means of macroscopic variables. A set of values of these variables
can be produced with many different microstates, and the number of
these microstates will define the probability $P$ of the macrostate
in question. At this point, an entropy $S$ and a temperature $T$
can be introduced, in purely formal terms, by means of the definitions:
\begin{equation}
S=k\ln{P},\label{eq: 25}
\end{equation}
where $k$ is the customary Boltzmann constant and 
\begin{equation}
\frac{dS}{dF}=-\frac{1}{T},\label{eq:26}
\end{equation}
 where $F$ is the energy that the system would liberate if all the
particles and fields which it is made of become real. By combining
the two relations, one has: 
\begin{equation}
P=\exp{\left(\frac{-F}{kT}\right)}.\label{eq: 27}
\end{equation}
Combining the expression for $T_{C}$ and the last one, one gets:
\begin{equation}
P=\exp{\left(\frac{-F\underline{x}_{0}}{\hbar c}\right)}=\exp{\left(\frac{-p^{0}\underline{x}_{0}}{\hbar}\right)}=\exp{\left(\frac{-\Sigma}{\hbar}\right)},\label{eq: 28}
\end{equation}
where $p^{0}=\frac{F}{c}$ and $\Sigma$ is the total action held
by the Universe \textquotedblleft before\textquotedblright{} the Big
Bang. It is interesting to note that the following relation exists
between the action and the entropy of the pre-Big Bang Universe:
\begin{equation}
\frac{\Sigma}{\hbar}=-\frac{S}{k}.\label{eq: 29}
\end{equation}
This can be seen by direct comparison with the definition of entropy
introduced before. In other words, $\Sigma$ is a negative entropy
or, one might say, a sort of information whose bit is $\hbar\ln{2}$.
From Eq. (\ref{eq: 29}), one has $\Sigma=\hbar\ln{P}.$ Thus, for
$P=\frac{1}{2}$ (binary choice), $\Sigma=\hbar\ln{2}$. In general,
a dimensionless amount of information $I=\frac{\Sigma}{\hbar\ln{2}}$
can be introduced. From the relation $\underline{x}_{0}\le\theta_{0}$,
which is valid in the \textquotedblleft pre-Big Bang\textquotedblright{}
era, if one puts $c\theta_{0}=2\pi\underline{R}$ one has $p^{0}\underline{x}_{0}\le2\pi p^{0}\underline{R}$,
i.e. $\Sigma\le2\pi F\underline{R}/c$. Thus, one can write
\begin{equation}
I\le\frac{2\pi F\underline{R}}{\hbar c\ln{2}}.\label{eq: 30}
\end{equation}
This is a form of the Bekenstein bound \cite{key-45}, which is valid
for the \textquotedblleft pre-Big Bang\textquotedblright{} phase.
Let us consider now the \textquotedblleft unfolding\textquotedblright{}
process of information, and operate a Wick rotation on Eq. (\ref{eq: 23}):
\begin{equation}
(\underline{x}_{0})^{2}-(\underline{x}_{1})^{2}-(\underline{x}_{2})^{2}-(\underline{x}_{3})^{2}-(\underline{x}_{5})^{2}=r^{2}.\label{eq: 31}
\end{equation}
The space expansion is described by the canonical extension of Eq.
(\ref{eq: 31}), that is
\begin{equation}
(\underline{x}_{0})^{2}-R^{2}(\tau)[(\underline{x}_{1})^{2}+(\underline{x}_{2})^{2}+(\underline{x}_{3})^{2}]+(\underline{x}_{5})^{2}=r^{2},\label{eq: 32}
\end{equation}
where $\tau$ is the cosmic time. An important difference with respect
to Friedmann cosmology is that, while it admits a multiplicity of possible
models, to be subsequently selected based on observation, the approach
described here leads to a single cosmological model. It corresponds
to the Friedmann model having null spatial curvature $(k=0)$ and
a positive cosmological term $\lambda=4/3t_{0}^{2}$. The reduction
of arbitrariness is a first mark of the power of this approach based
on group theory. Fixing the cosmological constant, one also fixes
a new natural constant $t_{0},$ which has the dimensions of time;
this time is related to the de Sitter radius $r$ through the relation
$r=ct_{0}$, where $c$ is the speed of light in a vacuum. At the
start of the expansion, i.e. $R(\tau)=0$, Eq. (\ref{eq: 32}) becomes
\begin{equation}
(\underline{x_{0}})^{2}+(\underline{x_{5}})^{2}=r^{2}.\label{eq: 33}
\end{equation}
Hence, if one assumes that the start of the expansion coincides with
the origin of $\underline{x}_{0}$, i.e. that the Big Bang occurs
on the equator $\underline{x}_{0}=0$ of the hypersphere (\ref{eq: 23}),
the value $\pm r$ is obtained for the variable $\underline{x}_{5}$.
In geometrical terms, this corresponds to a point-like Big Bang associated
with a point on the equator of the 5-sphere. However, the $\underline{x}_{5}$-axis
can be rotated on this equator giving rise to $\infty^{3}$ different
(and equivalent) intersections. Thus, one has $\infty^{3}$ different
(and equivalent) Big Bangs or, to be more precise, $\infty^{3}$ different
(and equivalent) views of the same Big Bang, which are pertinent to
distinct fundamental (inertial) observers. In individual observer's
coordinates, the metric is consistent with Eq. (\ref{eq: 32}) and
therefore all the observers see a Universe in expansion. At a certain
value of cosmic time $\tau$, all the observers see the Universe under
the same conditions. Thus, the cosmological principle works provided
that the conditions of matter on the equator $\underline{x}_{0}=0$
are homogeneous. The dimensionless vacuum starting from which the
Big Bang develops is therefore substituted, in this approach, by a
pre-existing space: the equator of the 5-sphere (\ref{eq: 23}). The
passage from the condition (\ref{eq: 23}) to the condition (\ref{eq: 32})
takes place at a critical value $\theta_{0}$ of the variable $\underline{x}_{0}/c.$
For that critical value, processes of quantum localization of elementary
particles on space\textendash time become possible. The emergence
of all the elementary particles on the space\textendash time domain
is the true essence of the Big Bang. Starting from this nucleation
by 5-sphere vacuum, the propagation of particles is described by wave-functions
in which coordinates satisfying the condition (\ref{eq: 32}) and
no longer the condition (\ref{eq: 23}) appear as an argument. The
\textquotedblleft archaic phase\textquotedblright{} governed by the
condition (\ref{eq: 23}) comes to an end and the actual history of
the Universe governed by the condition (\ref{eq: 32}) begins. And
time \textquotedblleft flows\textquotedblright . It must be noted
that the contraction resulting from the scale distance operates on
the private space\textendash times of the individual fundamental observers,
not on the public space\textendash time, which remains unchanged.
As one approaches the Big Bang proceeding backwards in cosmic time,
the private contemporaneousness space of each observer contracts in
one point; but the uncontracted public space will be identical for
all observers. Apart from fluctuations, the final mass-energy density
will be the same everywhere and will be equal to the ratio between
$F$ (the energy released in the transition) and the volume of the
section $\underline{x}_{0}=c\theta_{0}$, which is finite. Thus, there
is never a singular density value; in other words, in public space\textendash time
the Big Bang is not truly a singularity. Therefore, the origin of
Universe (and time) is a nucleation process implying a passage from
information to energy. Given the initial homogeneity, all the fundamental
observers will see the same physical cosmic conditions, despite the
absence of causal correlations between their respective positions.
Two difficulties with the standard model are bypassed in this way,
i.e. the justification of the initial homogeneity (which is here the
natural aspect of a pre-vacuum) and the appearance of a singularity.
Space\textendash time isotropy and homogeneity are the consequences
of the decay of an isotropic and homogeneous archaic (pre-)vacuum;
this line of reasoning agrees with some prominent features of other
contemporary approaches \cite{key-46}. Finally, as for cosmology,
it is interesting to notice that the wave function in archaic space
(\ref{eq: 23}), which is \cite{key-38,key-39} 
\begin{equation}
\Psi=\Psi_{0}\exp{\left[\pm i\frac{\sqrt{2mE}}{\hbar}x-\frac{E}{kT}\right]},\label{eq: 34}
\end{equation}
after the \textquotedblleft Big Bang\textquotedblright{} becomes \cite{key-38,key-39} 
\begin{equation}
\Psi=\Psi_{0}\exp{\left[\pm i\frac{\sqrt{2mE}}{\hbar}x-i\frac{Et}{\hbar}\right]}.\label{eq: 35}
\end{equation}
This solution is very similar to that of Hartle\textendash Hawking
Universe without boundaries \cite{key-46}, but also to the one proposed
by Bohm et al. for the processes of \textquotedblleft spontaneous
localization\textquotedblright{} \cite{key-47}. If one assumes as
valid the Born relation in the archaic phase, one can easily see the
meaning of the archaic wave-function. As a matter of fact, if one
puts 
\begin{equation}
P=\int dv\|\Psi\|^{2}=\exp{\left(-\frac{2E}{kT}\right)},\label{eq: 36}
\end{equation}
one sees that the probability of existence of an energy particle $E$
is not conserved at varying of temperature because the production
of a particle implies a clear exchange of energy with the thermal
bath: archaic QM is a form of thermostatics.

\section{What can cosmology teach on the interpretation of quantum mechanics?}

After the short examination of de Sitter Projective Holography in
previous Section, in this Section we bring back to the HH wave function
and to the oscillator in Eq. (\ref{eq: 12}). It is indeed argued
that the formal and physical meaning of the oscillator which works
with imaginary time can be clarified in the framework of de Sitter
Holography. From the geometrical point of view, in de Sitter S4 hypersphere
(such a hyphesphere can be also S5 if one assumes a 4-dimensional
hyphesphere which is embedded in a 5-dimensional space), space and
time are mixed. In fact, in order to have a space-time representation,
one needs to consider the projective representation of Figure 4.1
and, in turn, to consider the De Sitter-Castelnuovo-Beltrami projective
line element and the tangential space P4 where the grouping of observers
localize the physical events, see Figure 5.1. One sees that every
tangential space P4 ``inherits'' the behaviors of S4, and, then,
all the grouping of observers comply with a common cosmological event
horizon which is constrained by the value of the cosmological constant
and by the flatness of space-time. From the physical point of view,
this Archaic Universe is populated only by virtual events. These behaviors
identify the de Sitter S4 hypersphere as being a non-local place,
and this agrees with the imaginary and curved time introduced by Hartle
and Hawking in \cite{key-1}. The important difference is that, in the
framework of de Sitter Projective Holography, the pre and post Big Bang
phases do not stay on the same level. Instead, it is the background
geometry, which works with imaginary time, which fixes (constrains)
the evolutive dynamics which work with real time. Thus, the connection
between the two phases is not dynamical like in the original work
of Hartle and Hawking \cite{key-1}. It is instead holographyc through
the Wick transformation and the de Sitter projection. Roughly speaking,
one argues that the archaic vacuum gives the shape to the quantum
vacuum through the boundary conditions while the evolution of the
Universe represents its projection. Hence, this is a double representation
of the same physical object (entity) which is analogous to Milne models
\cite{key-48}. The Milne double description has been indeed recently
used in hadronic frameworks paying specific attention to de Sitter
Universe \cite{key-49,key-50}.

Let us recall that the peculiar problem of the Wheeler-DeWitt equation
is that the boundary conditions are non computable \cite{key-51}.
Therefore, the role of the double description is to fix the boundary
conditions in the hypersphere. Hence, we argue that the oscillator
in Eq. (\ref{eq: 12}) drives the nucleation in vacuum in agreement
with the general relations (\ref{eq: 34}) and (\ref{eq: 35}). One
can also try to understand the situation through the more traditional
and dynamical language of the tunnelling mechanism. In fact, particle
creation from quantum fluctuations is today studied in an elegant
and largely used way through tunnelling, starting from black hole
radiance \cite{key-52,key-53}. In a tunnelling framework, the whole
surface of the S4 hypersphere of Eq. (\ref{eq: 23}) drives the nucleation
for the value of the time precursor given by Eq. (\ref{eq: 24}).
The situation can be clarified in a better way through the holographic
language: the Universe of pure information of Eq. (\ref{eq: 30})
\textquotedblleft updates itself\textquotedblright{} becoming an observable
universe of real events through a ``turning around of time''. The
traditional picture of the Big Bang in terms of ``thermodynamics
balloon\textquotedblright{} results strongly transformed, as well
as the wave function of the Universe. Within the new scenario, the
Big Bang shows itself as the localization of particles in space-time,
that one labels as R Processes. This issue has been analysed in detail
in \cite{key-54,key-55}. Here we focus only on the cosmological framework.
We will consider the primordial R Processes, i.e. the first localizations
after the Big Bang (in other words, if one uses a cosmic clock, which here
corresponds with the group of privileged observers which are fixed by the de 
Sitter structure, one refers to the first instants after the Big Bang). 
From a formal point of view, it is possible to think about such R Processes 
as being the collapse of a Universe wave function having constant amplitude 
over the entire Universe (owing to the homogeneity resulting from the thermostatic 
behavior of the universal information reservoir contained in the Archaic Universe)
and a phase given by the action (\ref{eq: 29}) to the wave function entering into 
the Big Bang \cite{key-37} 
\begin{equation}
\Psi=\Psi_{0}\exp{\frac{-i\Sigma}{\hslash}}.\label{eq: 37}
\end{equation}
This wave function can also be a mere product of wave functions of
independent particles, which has not been symmetrized in any way,
despite it is possible to hypothesize the existence of a strong, non-local
pre-big bang correlation in terms of entanglement. If one considers
the relations (\ref{eq: 28}), (\ref{eq: 29}) and the Bekenstein-like relation 
(\ref{eq: 30}), one finds a similarity with the HH solution, despite there is 
some difference. For example, it results independent from considerations about different theories
of quantum gravity. This is because we chose to consider the chronon
as being a time scale well different from the Planck time. The relation
between these two time scales will be discussed in the Section of
the Conclusion Remarks. Here we note that in Eq. (\ref{eq: 34}) there
is an exponential dependence on the critical temperature which
is inverse with respect to exponential dependence on the pre-cosmic
state probability $x_{0}$. This is something like an \textquotedbl{}evanescent
wave\textquotedbl{} and suggests an analogy between the conversion $P \rightarrow \Psi$, and the tunnelling effect.
In fact, the Universe \textquotedblleft before\textquotedblright{}
the Big Bang seems a system that is classically stable which becomes
quantum-mechanically unstable generating the nucleation. Thus, it
is natural to suspect tunnelling and Eq. (36) is interpreted in terms of probability of tunnelling.
By using Eq. (\ref{eq: 29}), one indeed sees that the step from
Eq. (\ref{eq: 34}) to Eq. (\ref{eq: 37}) implies 
\begin{equation}
\left(\pm i\frac{\sqrt{2mE}}{\hbar}x-\frac{E}{kT}\right)\rightarrow\frac{iS}{k},\label{eq: (38)}
\end{equation}
while the step from Eq. (\ref{eq: 37}) to Eq. (\ref{eq: 35}) implies
\begin{equation}
\frac{iS}{k}\rightarrow\pm\left(i\frac{\sqrt{2mE}}{\hbar}x-i\frac{Et}{\hbar}\right).\label{eq: 39}
\end{equation}
Thus, one can interpret the step from Eq. (\ref{eq: 34}) to Eq. (\ref{eq: 35})
in terms of a variation of entropy 
\begin{equation}
i\triangle S=-ik\frac{Et}{\hbar}+\frac{E}{T}.\label{eq: 40}
\end{equation}
Again, we find an analogy with black hole radiance, where the tunnelling
of particles is drived by the change in the Bekenstein-Hawking entropy
of the black hole \cite{key-52}.
In any case, what has been obtained is an ``event based'' description
of the nucleation process. What has been loss in terms of traditional
descriptions of the wave function has been recovered without the usual
ambiguities connected to the mere juxtaposition between classical
and quantum models through the mixing of geometries which has been
discussed above. In the current approach the analysis is indeed focused
on the holographic correspondence between information and its localization
in time.
\begin{figure}
\includegraphics[scale=0.2]{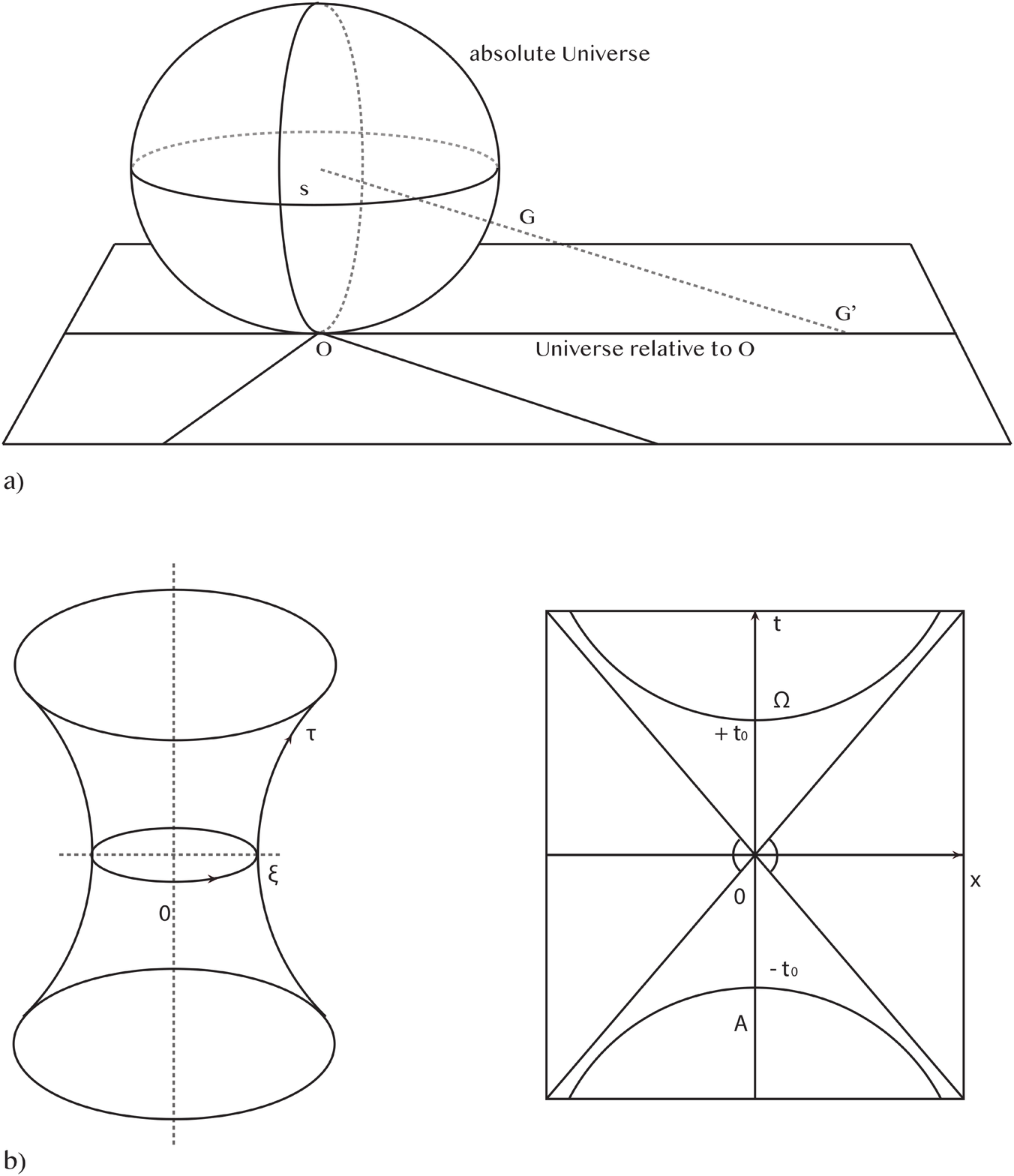}\caption{De Sitter and Beltrami Castelnuovo Representation: a) de Sitter: The hyperbolic  hyperboloid after the application of Wick rotation to S4; b) The P4 Beltrami-Castelnuovo Representation.}
\end{figure}
Another overturning is obtained in the relationship between cosmology
and QM interpretation. We sketched above that the absence of the observer
in cosmology  generates very peculiar interpretative issues, which
concern physical laws, boundary conditions and nucleation. On the
other hand, beyond the shortage of the current interpretations, here
one finds also the possibility that it could be the same quantum cosmology
to suggest a new class of interpretative strategies. This is currently
an active research field, which is for example connected to scenarios
of eternal inflation \cite{key-29}. In the framework of Projective Holography, the
constrains on the pre-space and on the localization time scale (the
chronon) and, in addition, the assumption of the validity of Born
Rule in the Archaic Universe, univocally select structure
and evolution of the Universe and enables the extension of this approach
to particles seen as ''Little Bangs''. 

\section{Conclusions. Holographies in comparison}

Starting from the Lagrangian of a homogeneous and isotropic Universe,
a relation between the HH cosmological wave function and a quantum
oscillator which works with imaginary time has been derived. This
permitted to return again to the problem of the physical meaning of
the HH solution and of the centrality of de Sitter Universe in theoretical
physics and in contemporary cosmology \cite{key-57}. 

In fact, starting from its original proposal \cite{key-1}, the HH
wave function received a lot of criticisms. Among such criticisms,
the most significant seems to concern the causual relationship between
an Euclidean state having imaginary time (without classical space-time)
and a Lorentzian metric having real time {[}58-60{]}. 

This kind of questions are present also in the Gamow-Vilenkin-like
approach \cite{key-61,key-60}, which is founded on the tunnelling
mechanism. In that case, the problem is the definition of the barrier.
On the other hand, this second big class of models benefited from
hypotheses and analogies with quantum field theory, particle physics
and black hole physics {[}62-67{]}.

Thus, it seems that, in its original proposal, the HH wave function
cannot escape the general fate of wave functions in quantum mechanics,
that is, from the year 1927 wave functions are only a formalism questing
a physical interpretation. Another requirement, which is even stronger
because of the unicity of the analysed system, is the elevation of
the boundary conditions to the level of physical law. The proposal
that we suggest in the present work is founded on the cosmological
implications of de Sitter relativity and on Projective Holography
\cite{key-35,key-68}. 

One recalls that in de Sitter Universe the cosmological constant is
no more a free parameter. It is instead structurally fixed through
the introduction of a global curvature which does not depend on gravitational
fields. One indeed assumes the hypersphere S4 as being a geometric
form of the quantum vacuum, which is, in turn, considered as universal
information reservoir or, in more physical terms, as being populated
only by virtual processes (imaginary time). This toy model of quantum
vacuum has been labelled ``Archaic Universe'' because its relationship
with observables is not chronological. It is logical instead. In fact,
the relationship between S4 and the standard post big bang scenario,
which works with real time and concerns an accelerated expansion of
the Universe, is obtained through Wick rotation and Beltrami-Castelnuovo
projection on a plane which is tangential to the hypersphere P4. One
notices that every group P4 ``inherits'' from S4 the same event
horizion, and so defines an equal substrate for all the observers.
Hence, it completely realizes Weyl's cosmological principle without
contradictions with the quantum realm \cite{key-69}.
Here De Sitter's eternal is only the hypersphere S4, while the quantities
which work with real time and refer to observables are attributed
to P4. There are neither physical singularities (but only apparent
ones due to the event horizon) nor contradiction (forcing) in the
transition between the two different geometries. We argue that the
most natural framework for the quantum oscillator (\ref{eq: 12})
should be the surface of S4. Due to time absence, this surface acts
as a kind of non local membrane. One among the coordinates of S4 has
the role of an archaic precursor of time and, due to the critical
temperature (\ref{eq: 24}), it starts to nucleate. In other words,
the virtual population starts to show itself in terms of real matter.
We argue that this should be the physical essence of the \textquotedblleft big
bang\textquotedblright{} and it seems that this kind of description
is not too much distant from a traditional description. In traditional
relativistic terms, one can say that the hypersphere represents a
global boundary condition concerning a big bang with an ``isotropic
singularity''. This is exactly what is requested by the Wheeler -
de Witt equation. The price to pay - which here seems to be favourable
- is a different interpretation of the cosmological wave function
in terms of product of the wave functions of the  nucleated particles.
In a oversimplified case, one considers non-symmetrized wave functions,
despite one can argue that the nucleated entities are strongly correlated
within a non-local framework. This is an \textquotedblleft event-based\textquotedblright{}
interpretation, which unifies cosmology and microphysics and is connected
with the popular \textquotedblleft timeless\textquotedblright{} approaches,
see \cite{key-18,key-30,key-70} and references within.

Generally speaking, every physical theory is realized through a formal
structure and an interpretation. The interpretation drives the meaning
of the formalism and ratifies the \textquotedblleft licit\textquotedblright{}
mathematical manipulations. In particular, it specifies the relationship
between theory and observations and experiments. Here, the \textquotedblleft event-based\textquotedblright{}
approach represents an extension of the standard QM, based on a time
scale for the localization. The meaning of the chronon must not be
considered as being a ``minimum time\textquotedblright . Instead,
one considers the chronon as being a time scale compatible with the
localization of baryon objects, that is, a kind of classical radius
analogous to Bohr's classical radius of the electron. This is what
one expects in a HH semi-classical approach which is very distant
from the Planck scale of quantum gravity. We also cite an interesting
relation between the two different scales \cite{key-30,key-55}, which
could be further analysed in future works.

In the framework of de Sitter relativity one finds de Sitter cosmological
horizon \cite{key-30,key-68,key-71,key-72}. The chronological distance
from every generic ``here'', ``now'' is a fundamental constant
of Nature, $t_{0},$ which is invariant in cosmic time, where $ct_{0}\approx10^{28}cm$
being $c$ the speed of light \cite{key-30}. One assumes that the
localization of an R process is associate with the formation of a de
Sitter micro-horizon centred in $O$ with a radius $c\theta_{0}\approx10^{-13}$cm,
where $O$ is generally delocalized in agreement with the wave function
which is entering/outing from the process \cite{key-30}. The constant
$\theta_{0}$ is independent of cosmic time \cite{key-30}. Thus,
also the ratio $\frac{t_{0}}{\theta_{0}}\approx10^{41}$ results independent
of cosmic time \cite{key-30}. This ratio expresses the number of
distinct time localizations which are accessible by the R process
within de Sitter cosmological horizon \cite{key-30}. Basically, $t_{0}$
represents the portion of straight timeline where an observer located
in the horizon's centre positions the R process, while the duration
of the R process is order $\theta_{0}.$ Thus, the portion of
straight timeline is split in about $10^{41}$ different ``cells''.
Each cell can stay in one among twt different states: ``on'' or
``off''. The time localization of a single R process corresponds
to a situation in which all the cells are ``off'' with a sole exception
of a particular cell which is ``on''. Configurations having more
that a sole cell which is ``on'' correspond to the localization
various different R process on the same straight timeline. If one
assumes that every cell is independent, one has a total of $\left(2^{10}\right)^{41}$
different configurations. Hence, the locational information which
is associate with the localization of $0,1,2,...,10^{41}$ R processes
amounts to $10^{41}$ bits, the binary logarithm of the number of
configurations. This is a kind of information which is codified on
the time axis within de Sitter horizon of the observer. The R processes
are indeed real interactions among real particles. During such interactions
an amount of action of order of the Planck constant $h$ is exchanged.
Then, in terms of phase space, the emerging of one among such processes
is equivalent to make ``on'' an elementary cell having a volume
of order $h^{3}$. The number of cells which have been made \textquotedblleft on\textquotedblright{}
in the phase space of a particular macroscopic physical system is
an estimation of the volume that such a system takes up in the same
phase space. In other words, it is an estimation of the entropy of
the system. Thus, one argues that the information on the localization
of the R processes should be connected to the entropy through the
uncertainty principle. Thus, it seems natural that one asks if Bekenstein
bound on entropy \cite{key-45} can be applied, in some way, to the
two mentioned horizons. Assuming that the information on the time
localization of R processes 
\begin{equation}
I\equiv\frac{t_{0}}{\theta_{0}}=10^{41}\,bits\label{eq: 41}
\end{equation}
is connected to the micro-horizon's area 
\begin{equation}
A=\left(c\theta_{0}\right)^{2}\approx10^{-26}\,cm^{2},\label{eq: 42}
\end{equation}
then, from the holographic relation \cite{key-30}
\begin{equation}
I=\frac{A}{4l^{2}},\label{eq: 43}
\end{equation}
one gets that the spatial extension $l$ of the ``cells'' which
are associated to a bit of informations results to be 
\begin{equation}
l\approx10^{-33}\,cm,\label{eq: 44}
\end{equation}
that is the Planck length! On one hand this is an intriguing, remarkable
issue. On the other hand it cannot be a coincidence that the Planck
scale appears in this way, i.e. as a consequence of the holographic
relation (\ref{eq: 43}) combined with the ``two horizons'' hypothesis
and with the finite behavior of the information $I.$ In addition,
as both $I$ and $t_{0}$ are connected with the cosmological constant
$\Lambda$ through the relation 
\begin{equation}
\Lambda=\frac{4}{3t_{0}^{2}},\label{eq: 45}
\end{equation}
Eq. (\ref{eq: 43}) substantially defines the Planck length in function
of the cosmological constant. In other words, one gets a global-local
relation, which is exactly what one expects from a vacuum theory interpreted
in terms of information universal reservoir. In this paper, we referred
to the connection between the oscillator's frequency and the cosmological
constant. From the above discussion, one can reasonably argue that
the holographic conjecture should be a property of the two time-like
horizons: the cosmological horizon ($t_{0}$) and the \textquotedblleft particle-like\textquotedblright{}
horizon ($\theta_{0}$). This holographic conjecture cannot be generalized
to other physical systems (with the exclusion of black holes, because
black holes have their proper event horizon). The information associate
with the time localization of a physical event (R process) has finite
behavior. The same happens for its spatial localization. This is because
an observer sees the extension of a transition in Beltrami time as
having a lower bound due to the \textquotedblleft particle-like\textquotedblright{}
horizon and a higher bound due to the cosmological horizon. This suggests
an interesting interpretation of the cosmological constant in terms
of ``index of spreading information''. In addition, the old matter
on the nature of the cosmological constant in terms of geometry vs
physics here is bypassed by the Milne-like interpretation. On one
hand, in S4 $\Lambda$ appears as a purely geometric ingredient, pertinent
to the structure of de Sitter space. On the other hand, in P4 , after
the nucleation process and the conversion from information to matter,
it is both licit and necessary interpreting the cosmological constant
in terms of fields and particles, in particular concerning the open
issues on Higgs field and Dark Matter {[}73 - 77{]}. The analysis
in this paper is limited to a semi-classical approach. Instead, the
process of nucleation should imply the emergence of both space and
time from a web of elementary events. This is a research field which
strongly involves quantum field theory, particle physics and the ``universal''
role of Higgs field . These issues could be the object of future works
of the authors.

We also recall that, after some purely quantum mechanical approach,
in the recent literature there are also models of emergent space-time
founded on non-locality {[}78 - 80{]}. In the scenario analysed in
this paper, non-locality means that the Universe's wave function could
nucleate events which are strongly correlated under a drastic, but
plausible, symmetrization hypothesis. The mathematical behaviors connected
to the maximum symmetry of de Sitter space always fascinated and fascinate
theorists, who, in the meanwhile, proposed cosmological and particle
physics models \cite{key-35}. A relevant matter is the development
of a consistent quantum field theory in de Sitter space-time. In this
framework, the most important problem concerns the stability of de
Sitter vacuum and the topological classifications of its decays {[}81
- 83{]}. In our opinion, Projective Holography can introduce some
clarification in the current debate, by enabling a clear distinction
between S4 and P4 \cite{key-84}. In that direction it could be possible
a better understanding of the most traditional holographic approaches
connected to Maldacena conjecture \cite{key-85}, in particular through
 the advent of a new centrality of de Sitter space-time with respect
to Anti-de Sitter space-time. In fact, one among the most important
problems in this research field consists in projecting an hologram
from quantum particle living in the infinite future. This strongly
obstructed various efforts in order to describe in holographic terms
the real space-time. It is possible to analyse this problem in a S4/P4
framework in order to get a brighter formulation, which seems more
practicable after the uplifting result in \cite{key-86} which transformed
two Anti-de Sitter spaces having a saddle shape in de Sitter spaces
having bowl shape. Finally, for the sake of completeness, we also
signal the recent CPT cosmological hypothesis in \cite{key-87}, which
arises from the issue that the S4 structure suggests two specular
worlds (most likely Anti-de Sitter spaces). De Sitter Projective Holography
shows a structure which is naturally holographic, where the 5D pre-space
works as a screen and its projections on the tangential plane P4 represent
the bulk, i.e. the expression of the real physical world. In Archaic
Holography, the localization and delocalization processes seem as
complementary world aspects, while the uncertainty principle represents
a ``door'' between two different levels of physical descriptions:
from particle physics to cosmology. Both of them are strongly unified
in a global-local connection. On the other hand, going beyond that
``door'' shows to be less dramatic than physicists were thinking
in past research works. In addition, this new scenario seems even
more understandable that previous classical scenarios. The relations
among various physical quantities like cosmological constant, chronons
and Planck length are strongly interconnected and, globally, they
indicate the finite behavior of information in Nature. 

\section{Acknowledgements }

I. Licata and C. Corda have been supported financially by the Research
Institute for Astronomy and Astrophysics of Maragha (RIAAM).
The authors thank Maria Vita Licata for checking the English language.

\end{document}